\documentclass[%
 twocolumn,
superscriptaddress,
 amsmath,amssymb,
 aps,
 prl,
floatfix,
]{revtex4-2}

\usepackage{kantlipsum}
\usepackage{xcolor}
\usepackage{graphicx}
\usepackage{dcolumn}
\usepackage{bm}
\usepackage[caption=false]{subfig}
\usepackage{orcidlink}

\usepackage{hyperref}
\hypersetup{pdfauthor=Fabbricatore,
    colorlinks, linktocpage=true, pdfstartpage=3, pdfstartview=FitV,
    breaklinks=true, pdfpagemode=UseNone, pageanchor=true, pdfpagemode=UseOutlines,
    plainpages=false, bookmarksnumbered, bookmarksopen=true, bookmarksopenlevel=1,
    hypertexnames=true, pdfhighlight=/O,
    urlcolor=orange, linkcolor=blue, citecolor=red, 
        }
        
\newcommand{\Cov}{\mathrm{Cov}}

\begin{document}

\preprint{APS/123-QED}

\title{Gradient dynamics in reinforcement learning}

\author{Riccardo Fabbricatore\,\orcidlink{0000-0002-6298-2885} }
\email{ri.fabbricatore@gmail.com}
\author{Vladimir V. Palyulin \,\orcidlink{0000-0002-3047-6937}}%
 \email{v.palyulin@gmail.com}
\affiliation{Skolkovo Institite of Science and Technology, 121205, Moscow, Russia
}%



\date{\today}

\begin{abstract}
Despite the success achieved by the analysis of supervised learning algorithms in the framework of statistical mechanics, reinforcement learning has remained largely untouched. Here we move towards closing the gap by analyzing the dynamics of the policy gradient algorithm. For a convex problem, we show that it obeys a drift-diffusion motion with coefficients tuned by learning rate. Furthermore, we propose a mapping between a non-convex reinforcement learning problem and a disordered system. This mapping enables us to show how the learning rate acts as an effective temperature and thus is capable of smoothing rough landscapes, corroborating what is displayed by the drift-diffusive description and paving the way for physics-inspired algorithmic optimization based on annealing procedures in disordered systems.
\end{abstract}

\maketitle


\section{Introduction}
Statistical mechanics is a powerful tool for understanding and constructing optimization algorithms. On one hand, disordered systems, such as spin glasses or polymers, prompted the development of new algorithms (simulated annealing \cite{Kirkpatrick671}, cluster algorithms \cite{wolff89}, hysteric optimization \cite{zarand02}). On the other hand, existing optimization algorithms have often been fruitfully analyzed in the statistical physics' framework, yielding knowledge about their behavior, phase transitions and possible improvement \cite{mezard02, franz02, hartmann03, lukasz17, montanari02}.

In recent years, the vast class of machine learning algorithms \cite{jordan2015machine} has enjoyed a great deal of attention. Neural networks \cite{goodfellow2016deep, aggarwal2018neural} are nowadays used to predict protein folding \cite{jumper2021highly}, search for exotic particles in high-energy colliders \cite{baldi2014searching}, predict phase transitions \cite{wetzel2017phase}, and in many other fields \cite{carleo2019machine}. At the same time, reinforcement learning \cite{sutton2018reinforcement, mnih2015human} has proven to be a valuable tool for finding optimal jet grooming strategies \cite{carrazza2019jet}, in the pursue of the conformal bootstrap program \cite{kantor2022conformal}, or in the engineering of smart active matter \cite{celani2017flow}. Nonetheless, numerous questions about the algorithms' functioning remain unanswered \cite{zdeborova2020understanding}.
Great progress has been made in the study of neural networks, the analogy between their highly non-convex loss function landscapes and the free energy landscape of disordered systems has been extensively studied \cite{gardner1988optimal,barkai1992broken,huang2014origin}.
It has been shown how the stochastic gradient descent algorithm \cite{robbins1951stochastic, bottou2010large} is prone to lead the network's weights towards a needed suboptimal, robust, and well-generalizing region \cite{baldassi2016unreasonable, feng2021inverse}.
However, all the results above are applicable to supervised learning problems, which can be mapped to disordered systems by interpreting the loss function as a Hamiltonian.

Despite their late successes, reinforcement learning algorithms have not yet received such analysis. This is perhaps due to the lack of a clear mapping between RL problems and disordered systems. We try to overcome this gap by studying a subset of reinforcement learning algorithms named policy gradients (PG) \cite{williams1992simple,sutton2000policy}. PG are the most universal training methods for reward-driven learning, they can be applied without additional knowledge of the agent's surrounding. Their main disadvantage is their tendency to converge to local maxima, thus learning a peculiar behavior, heavily dependent on the initial parameters. Nonetheless, PG-based algorithms were applied with a tremendous success in areas such as robotics \cite{andrychowicz2020learning}, natural language processing \cite{paulus2017deep}, and games \cite{berner2019dota}. A proper understanding of the reasons of this success is still an open question. We obtain a description for the learning process in a convex landscape in terms of drift-diffusion dynamics. By mapping a non-convex RL setting to a spin glass at a finite temperature, we are able to explain the effect of hyperparameters on the learning success thanks to a mean-field analysis. As it turns out, the learning rate is coupled to the temperature and, thus, its variation allows one to perform an annealing.


\section{The reinforcement learning framework}
The typical reinforcement learning setting, the so-called \textit{Markov decision process} \cite{bellman57}, consists of an agent acting in an environment with the purpose of maximizing a given utility function. The agent bases its decisions on the environmental \textit{state} $s\in \mathcal{S}$, choosing an \textit{action} $a\in \mathcal{A}$, according to its \textit{policy} $\pi(a|s)$. Subsequently, it receives a feedback from the environment in terms of a \textit{reward} $R \in \mathbb{R}$ and the state of the environment changes to a new one $s\rightarrow s'$. The reward is generated from a distribution conditioned to the state and the chosen action $q(r|s,a)$ and the transition between states is governed by the probability density $p(s'|s,a)$.
From this new state, a new action can be taken, generating again a new reward and a new state-transition.
The sequence of rewards obtained through this iteration is the agent's maximization goal. The central evaluated quantity is the \textit{return}: $G = \sum_{t=0}^\infty R_t \gamma^{t}$, i.e. the sum of the obtained reward sequence discounted by a factor $\gamma$, $0 \leq \gamma < 1$, which tunes the importance of memory.
Note that we used capital letters for $R$ and $G$ because they are, in general, stochastic variables.
The utility function of the agent is the average return: $Q_\pi(s,a) = E_{\pi,p,q} [G | S_0=s,A_0=a]$.
Denoting the distribution of initial states $\rho_0(s)$, the expected return of the policy $\pi$ reads:
\begin{equation}
J_\pi = \sum_s \rho_0(s) \sum_a \pi(a|s) Q_\pi(s, a) .
\label{eq:return}
\end{equation}

Reinforcement learning aims to efficiently find a policy $\pi$ that maximizes $J_\pi$. In general, the agent does not know the rules that govern the environment (e.g. $p$ and $q$), and it must build its strategy based on the information that it acquires while learning.

In this Letter we analyze \textit{policy gradient} algorithm \cite{sutton2018reinforcement}.
It exploits the well-known idea of gradient ascent to find the maximum of the return function (\ref{eq:return}). 
In this case the policy $\pi(a|s,\boldsymbol{\theta})$ is parametrized with a $d$-dimensional set of numbers $\boldsymbol{\theta} = \{ \theta_1, \ldots, \theta_d\}$.
The gradient ascent consists in updating these parameters in the direction of the steepest ascent of the average return (\ref{eq:return}). At state $s$ and for action $a$ it can be proven to be $\partial_{\boldsymbol{\theta}} J(\boldsymbol{\theta}) = E_{\pi, q, p} \left[ Q_\pi(s, a) \partial_{\boldsymbol{\theta}}\log\pi(a|s,\boldsymbol{\theta}) \right]$.
However, since the agent does not know how to compute this average (it does not know $p$ and $q$, as well as the utility function), it has to rely on an estimate of this gradient. One solution is to use the quantity $(G(s,a) - h(s)) \partial_{\boldsymbol{\theta}} \log\pi(a|s,\boldsymbol{\theta})$, where $G(s,a)$ is an estimate of the quality function, and $h(s)$ is an arbitrary action-independent function called \textit{baseline}.
At each time step $t$, the new parameters $\boldsymbol{\theta}_{(t+1)}$ will be derived from the current ones $\boldsymbol{\theta}_{(t)}$ by adding the gradient, multiplied by a coefficient $\alpha$, called \textit{learning rate}. To render the procedure invariant from the policy parametrization, one can fix the Kullback-Leibler divergence $D(\pi_{t+1} || \pi_t)$ at all steps, therefore obtaining the so-called \textit{natural policy gradient} \cite{kakade2001natural, bhatnagar2009natural}:
\begin{equation}
\begin{split}
\boldsymbol{\theta}_{(t+1)}  = \boldsymbol{\theta}_{(t)} + \alpha \; F_{(t)}^{-1} \; & \left( G(s_{(t)}, a_{(t)}) - h(s_{(t)}) \right) \\ 
&\times \partial_{\boldsymbol{\theta}}\log\pi(a_{(t)}|s_{(t)},\boldsymbol{\theta}_{(t)}),
\end{split}
\label{eq:natural_PG}
\end{equation}
where
\begin{equation}
(F)_{ij} = E_\pi \left[ \partial_{\theta_i} \log \pi(a|s,\boldsymbol{\theta}) \partial_{\theta_j} \log \pi_(a|s,\boldsymbol{\theta}) \right].
\end{equation}
The matrix $F$ is the \textit{Fisher information metric} of the policy for the parameters $\boldsymbol{\theta}$ \cite{amari2016information}.
There are several ways to choose $G(s,a)$, defining different types of policy gradient algorithms.
One straightforward possibility is to compute the future return by sampling the rewards for the next step of the process at fixed policy. 
This procedure is called \textit{reinforce policy gradient} \cite{williams1992simple}.

\section{Diffusion approximation for one-dimensional k-armed bandit}
We will begin our analysis by studying a case in which a single agent can use $k$ actions in an environment composed of only one state. Such a problem is known in literature as \textit{k-armed bandit} \cite{lattimore2020bandit} since it is analogous to a slot machine with $k$ arms, for which the player must infer which arms give better rewards, whilst trying to maximize his win. 
We will start with a scenario with only two possible actions: $\mathcal{A} = \{1,2\}$.
Since the gradient is not affected by the particular parametrization choice, we will use the convenient softmax function:
\begin{equation}
\pi(1|\theta) = x(\theta) = \frac{1}{1+e^{-\theta}}, \quad \pi(2|\theta) = 1-x(\theta).
\label{eq:param}
\end{equation}
At every step $t$, the agent will choose actions $1$ and $2$ with probabilities $x(t) \equiv x(\theta(t))$ and $1-x(t)$, respectively. This will yield the total average return (\ref{eq:return}) for $\gamma = 0$:
\begin{equation}
    J(\theta(t)) = x(t) R_1 +(1-x(t))R_2,
\end{equation}
where $R_a$ represent the stochastic reward extracted from its corresponding distribution $R_a \sim q_a=\mathcal{N}(r_a,\sigma_a)$.
The bandit setting allows us to choose a zero discount factor $\gamma=0$ without losing generality since the best policy is independent of it and we will keep this through the rest of this Letter.

Our aim is to obtain an effective stochastic description of the temporal evolution of the learning process, i.e. of the trajectory of the policy $x(t)$. In supervised learning, the effective noise of stochastic gradient descent is often modeled by heavy-tailed distributions \cite{gurbuzbalaban2021heavy,xie2020diffusion}. In our case, since the stochasticity is induced by uncorrelated Gaussian fluctuations in the rewards, we can describe the process in terms of a Langevin equation:
 \begin{equation}\label{eq:lang}
    \frac{dx}{dt} = u(x) + \sqrt{2D(x)}\cdot \eta_t,
    \end{equation}
    where $\eta_t$ is white Gaussian noise with zero mean and correlation $E_t[\eta_\tau\eta_{\tau'}] = \delta(\tau-\tau')$.
To this end, we expand the policy for small $\alpha$ by Taylor series:
\begin{equation}
    d x(t)= \left.\frac{d x}{d\theta} \right|_{\theta=\theta_{(t)}} d\theta_{(t)}+ \left. \frac{1}{2}\frac{d^2 x}{d\theta^2}\right|_{\theta=\theta_{(t)}} d\theta_{(t)}^2 +o(\alpha^2).
\end{equation}
Substituting the parameter update (\ref{eq:natural_PG}) in this expression, and computing the derivatives of (\ref{eq:param}), we obtain the policy increments. 
The drift and the diffusion terms are given by the average and the variance of these increments, $u(x) = E_{t} [\dot{x}(t) | x(t)]$, and $D(x) = \text{Var}_{t} [\dot{x}(t) | x(t)]/2$. We refer the reader to the Supplemental Material for a thorough derivation of these terms, while reporting here only their final form obtained by expanding up to the second order in $\alpha$:
\begin{equation}
\begin{aligned}
 u(x) =& \; \underbrace{ \alpha x (1-x) (r_1-r_2)}_{\text{Selection}} + \underbrace{ \frac{\alpha^2}{2} (1-2x) \; m }_{\text{Mutations}},
\\
 D(x) = & \underbrace{\frac{\alpha^2}{2} x (1-x)d_1 +\frac{\alpha^4}{4}(1-2x)^2d_2}_{\text{Random genetic drift}} .
\end{aligned}
\label{eq:coeff_NPG_diff}
\end{equation}
The three coefficients $m$, $d_1$ and $d_2$ are positive and depend on the reward variances as well as the policy, the average rewards, and the baseline:
\begin{equation}
\begin{aligned}
m=& (1-x) \left(\sigma_{R1}^2 +  l_1^2 \right) + x \left(\sigma_{R2}^2 + l_2^2 \right),
\\
d_1 =& (1-x) \sigma_{R1}^2 + x \sigma_{R2}^2 +  \left[ (1-x)l_1 + x l_2 \right]^2,
\\
d_2=& (1-x)^2\frac{(3-x)c_1^2-2l_1^4}{x} 
+x^2 \frac{(2+x)c_2^2-2l_2^4}{1-x} ,
\end{aligned}
\label{eq:mut_drift_funct}
\end{equation}
where $c_a=\sigma_a^2+l_a^2$ and $l_a=r_a-h$.

It is interesting to highlight the similarity with an evolving population of competing species/genotypes, described by the Kimura equation \cite{kimura1964diffusion,baake2000biological}:
\begin{equation}
\begin{aligned}
& u_{K}(x) =  \underbrace{ x (1-x) (f_1-f_2)}_{\text{Selection}} \underbrace{ - \mu_{12} x + \mu_{21}(1-x)}_{\text{Mutations}},
\\
& D_{K}(x) = \underbrace{\frac{1}{2 N} x (1-x)}_{\text{Random genetic drift}} ,
\end{aligned}
\label{eq:kimura}
\end{equation}
where $f_i$ is the fitness of the genotype $i$, $\mu_{ij}$ is the mutation rate from genotype $i$ to $j$, and $N$ is the population size.
The mapping can be done by identifying genotypes with the actions and the policy of each action with the genotype frequency. 
In contrast to our expansion, the Kimura equation is obtained by manually adding the evolutionary forces: \textit{selection}, \textit{mutation} and \textit{random genetic drift}. Our derivation can perhaps be considered more natural and clearly shows the symmetry between the deterministic and stochastic forces, adding a term proportional to $(1-2x)$ in the diffusion coefficient.
\begin{figure}
\includegraphics[width=1.0\columnwidth]{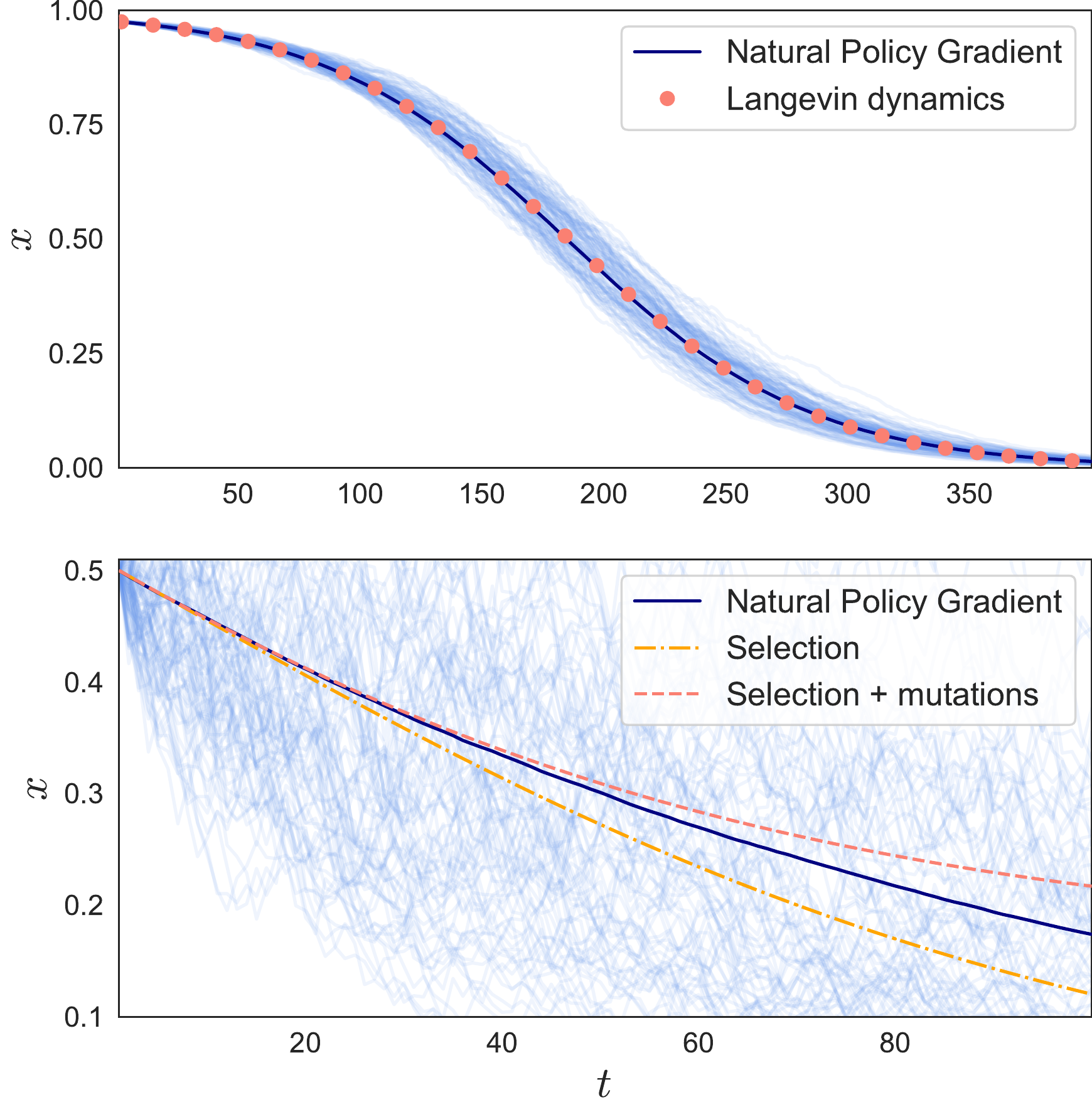}
\caption{Top: $10^2$ lightly shaded trajectories of the action probability $x$ generated by a natural policy gradient for the 2-armed bandit, along with their mean, compared to the Langevin dynamics (\ref{eq:coeff_NPG_diff}). The rewards are distributed as $\mathcal{N}_{1(2)}(r=\pm1,\sigma=1)$, while the learning rate is $\alpha=0.01$, and the initial policy is close to the worst one $x_0=0.975$. Bottom: The contributions of mutation and selection on the average Langevin dynamics near the boundaries, compared to the natural policy gradient. Rewards are distributed as $\mathcal{N}_{1(2)}(r=\pm1,\sigma=9)$, $\alpha=0.01$, $x_0=0.5$.}
\label{fig:exp_sim_slow_learning}
\end{figure}

It is easy now to grasp how this dynamics evolves and how it is affected by the algorithm's parameters. Figure \ref{fig:exp_sim_slow_learning} shows the effects of the drift coefficient on the gradient dynamics. The two terms correspond to natural selection and mutations, and can be tuned with the learning rate.
For a large learning rate, the policy is pushed away from pure strategies, i.e. vertices of the probability simplex. Conversely, for small learning rates, the policy tends to converge to the best action. The intrinsic stochasticity of the algorithm appears in the diffusion coefficient (\ref{eq:coeff_NPG_diff}): small learning rates confine stochasticity to the bulk of the strategy simplex ($x\approx 1/2$), while higher rates will generate higher fluctuations in the vicinity of pure strategies, as shown in appendix II.

These insights can be used to improve the dynamics' convergence by treating the learning rate as a dynamical variable, which can be tuned according to a time schedule \cite{darken1992gradient}. The approximation in terms of an It\^{o} stochastic equation allows us to use It\^{o}'s lemma to derive the optimal scheduling of the learning rate. This turns out to be $\alpha(t)\propto 1/\sqrt{t}$, which is consistent with the results for the so-called Exp3 algorithm \cite{lattimore2020bandit}, all details of the derivation can be found in appendix I.

All the obtained results can be easily generalized for the case in which the agent has $k$ possible actions and their probabilities follow a $k$-dimensional drift-diffusion motion:
\begin{equation}
 d\pi_a = u_a(\pi)dt + \sum_{ab}^N\sigma_{ab}(\pi) dW_b \quad,\label{sup:multidim}
\end{equation}
expressed here in the It\^{o} form. The resulting coefficients for this motion are
\begin{equation}
\begin{aligned}
     u_a = & \alpha\pi_a \Bigg( r_a-\sum_b r_b\pi_b\Bigg) + \\
     & \frac{\alpha^2}{2}\Bigg(\sigma_a^2 (1-\pi_a)(1-2\pi_a) -\sum_{b\neq a} \sigma^2_b(1-2\pi_b)\pi_a \Bigg),\\
 &\\
       D_{ab} = & \frac{\alpha^2}{8} \pi_a\pi_b\Bigg(\delta_{ab}\frac{\sigma_a^2}{\pi_a} + \sum_{c\neq a,b} \pi_c \sigma_c^2
       \\
       &  \qquad \qquad -(1-\pi_a)\sigma_a^2 - (1-\pi_b)\sigma_b^2  \Bigg).
\end{aligned}
\label{eq:coef_simply}
\end{equation}
They drive the trajectory towards the best action by a so-called replicator dynamics \cite{schulster83} proportional to $\alpha$, and away from pure strategies by the mutation term proportional to $\alpha^2$. In addition, the diffusion term scatters the trajectory proportionally to the rewards' variances. A thorough derivation of these results is reported in the Supplemental Material.

\section{p-dimensional k-armed bandit}
\begin{figure}
\includegraphics[width=.9\columnwidth]{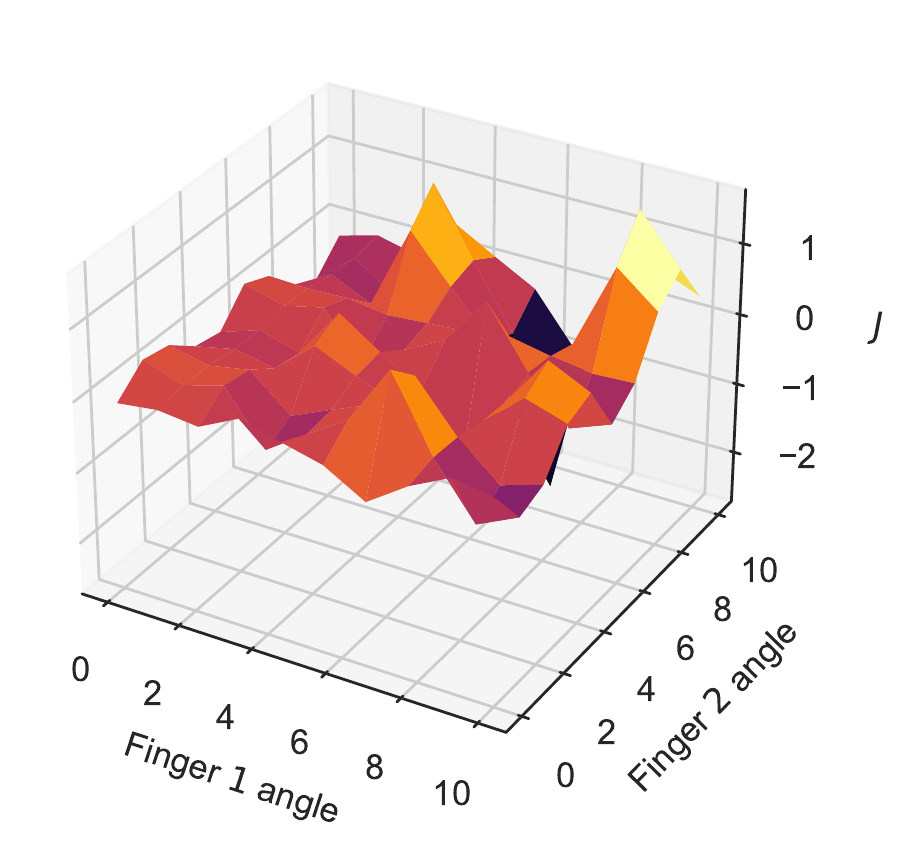}
\caption{An example of the return (energy) landscape of a robotic hand bending two fingers. Each finger can bend to $11$ different angles, the return $J$ is a function of the overall configuration.}
\label{fig:hand_landscape}
\end{figure}

The $k$-armed bandit can be viewed as a special case of a more general model in which the return is expressed as 
\begin{equation}\label{eq:pdim}
    J = \sum^K_{i_1,i_2,\ldots,i_p=1}R_{i_1 i_2 \ldots i_p} \pi_{i_1}\cdot\pi_{i_2}\cdot\ldots\cdot\pi_{i_p},
\end{equation}
where $\sum_{i=1}^K \pi_i = 1, \; \pi_i\geq 0 \; \forall i\in\{i_1,\dots,i_p\}$. Each probability distribution $\pi_i$ is defined over a distinct set of $K$ actions. All $p$ such sets are independent. This picture can be viewed simply as a factorization of the overall distribution $\pi=\prod_i \pi_i$. It arises naturally when one deals with an agent performing a set of actions at each time step and the task is to optimize the resulting overall behavior. For instance, robotics deals with a multitude of artificial joints flexed simultaneously \cite{andrychowicz2020learning, todorov18}, producing a highly non-convex cost landscape, as portrayed in Fig. \ref{fig:hand_landscape}. Furthermore, this model describes $p$ interacting agents, each performing independently their set of $K$ actions \cite{littman1994markov}. The reward coefficients of each agent $R_{i_1\ldots i_p}$ could be different in this case, but for equal constant coefficients, this is a generalization of the random replicant model \cite{diederich89, opper92, biscari95}.  Another useful interpretation arises when an agent is performing a sequence of actions in a state-changing environment so that for each state $s_t$, $\pi_t$ is the policy over the set of its $K$ actions. The ordered set $(\pi_1,\pi_2,\dots,\pi_p)$ then corresponds to the sequence of policies undertaken.

What is remarkable about this model is that now we have a clear way to map a reinforcement learning problem to a disordered system. This can be achieved by taking the instantaneous rewards to be normally distributed around their mean values $\mathcal{N}(\overline{R}_{i_1 i_2 \ldots i_p},\sigma_{i_1 i_2 \ldots i_p})$, and considering the system described by the Hamiltonian $H\equiv -\overline{J}$, obtained substituting mean rewards in (\ref{eq:pdim}). Its temperature $T(\sigma)$ is defined by the specific learning algorithm, and for a policy gradient is proportional to the diffusion coefficient of the Langevin dynamics (\ref{eq:lang}).

 PG dynamics is described by a system of $p$ multidimensional Langevin equations, navigating through the rough landscape of (\ref{eq:pdim}). To evaluate the effect of the learning rate on this motion, we will shift our perspective from the probabilities $\pi$ to the parameters $\theta$. The latter form a basis defined by
\begin{equation}
    d\boldsymbol{\theta} \propto \alpha \nabla\ln \pi = \nabla \ln \phi, \qquad \phi=\pi^\alpha.
\end{equation}
In other words, we move from a picture in which the learning rate is affecting the parameters' change to the one where the learning rate is affecting the slope of the probability manifold. We can define the following Hamiltonian for this new landscape,
\begin{equation}\label{eq:newham}
    H=-\sum_{i_1,i_2,\ldots,i_p}^K \overline{R}_{i_1i_2\ldots i_p} \phi^{1/\alpha}_{i_1}\phi^{1/\alpha}_{i_2}\ldots \phi^{1/\alpha}_{i_p}.
\end{equation}
We take $K$ to be large and mean rewards to be self-averaging, i.e. distributed as $\overline{R}\sim\mathcal{N}(0,\sigma)$ with $\sigma^2\sim 1/K$. This allows us to conveniently exploit methods of mean-field theory to analyze this landscape \cite{mezard1987spin}. Its average partition function over the variables $\pi$ will look similar to the partition function of the spherical $p$-spin \cite{kirkpatrick1987p,crisanti1992spherical} with planar rather than spherical constraints:
%

\begin{equation}\label{eq:part1}
\begin{split}
\langle Z\rangle &= \int_0^\infty\prod_{i=1}^K d^p\pi_i \, \delta^p(\sum_i\pi_i - K)  
\\
&\times \int_{-\infty}^{+\infty} \prod_{i_1,\ldots,i_p} d\overline{R}_{i_1\ldots i_p} \\
\\
& \times \exp{\left[-\overline{R}^2_{i_1\ldots i_p}K^p +\beta \overline{R}_{i_1\ldots i_p}\pi_{i_1}\ldots\pi_{i_p}\right]},
\end{split}
\end{equation}
where $\beta=1/T$. This expression can be rendered tractable by the replica trick $\langle\ln Z\rangle = \lim_{n\to 0} \frac{1}{n} \ln \langle Z^n\rangle$ in order to compute its mean value.
\begin{equation}\label{eq:pspin}
    \langle Z^n\rangle = \int D\pi \exp{\left[\frac{\beta^2}{4K^{p-1}}\sum^n_{a,b}\left(\sum_i^K\pi^a_i\pi^b_i\right)^p\right]},
\end{equation}
where $\int D\pi$ is a shorthand for the measure $\prod_{a=1}^n\prod_{i=1}^K d\pi_i^a \, \delta(\sum_j\pi_j^a - K)$.
Introducing $Q_{ab}=\sum_i\pi_i^a\pi_i^b$ by inserting the identity $1=\int\delta(Q_{ab} - \sum_i\pi_i^a\pi_i^b)\,dQ_{ab}$, and changing to Fourier representations for all delta functions, we obtain
\begin{equation}
\begin{split}
    \overline{Z^n} & = \int \prod_{a,b}^{n}\prod_{i}^KdQ_{ab}d\Lambda_{ab}d\xi^a d\pi^a_i \cdot \\
    & \cdot \exp\left[   \frac{\beta^2 K}{4} \sum_{ab}Q^p_{ab} + K\sum_{ab}Q_{ab}\Lambda_{ab} \right.\\
  &  \left. - \sum_i\sum_{ab}\Lambda_{ab}\pi^a_i\pi^b_i -\sum_{ia}\xi^a\pi^a_i + K\sum_a\xi_a \right].
\end{split}
\end{equation}
For large $K \rightarrow \infty$, the integral is dominated by the saddle point of the exponent's argument, thus the free energy can be recovered by solving a system of equations.

In the neighborhood of a pure strategy (where $\pi_a\approx 1, \; \pi_b\approx0 \; \forall \; b\neq a$), the partition function for the Hamiltonian (\ref{eq:newham}) can be recovered from Eq. (\ref{eq:pspin}) by substituting $p\rightarrow p/\alpha$. This will affect the saddle point equation containing the temperature
\begin{equation}
    0=\frac{p}{4T^2\alpha}Q_{ab}^{\frac{p}{\alpha}-1} + \Lambda_{ab}
\end{equation}
in a fundamental way: It will get modified by $T \rightarrow \sqrt{\alpha} T$. Thus, $\sqrt{\alpha}$ acts as an effective temperature that modifies the shape of the free energy landscape. 


\section{Discussion}
Our analysis sheds light on the ability of policy gradient to overcome obstacles in complex reward landscapes. It appears that the dynamics of policies under PG follows a drift-diffusion motion with parameters strongly influenced by the learning rate. Higher values of the latter allow the policy to scatter and overcome obstacles. This picture is corroborated by our mean-field analysis of the free energy landscape for a complex reward scenario, with multiple local minima. The learning rate appears to act as an effective temperature smoothing the free energy landscape. It follows that scheduling of this parameter is essential to ensure the convergence to high value maxima. Furthermore, it follows that this scheduling corresponds to the physical process of annealing. This paves the road to a plethora of physics-inspired optimizations (as proposed, for instance, in \cite{zarand02, houdayer99, mobius97}) to PG algorithms.
\begin{figure}
\includegraphics[width=.49\columnwidth]{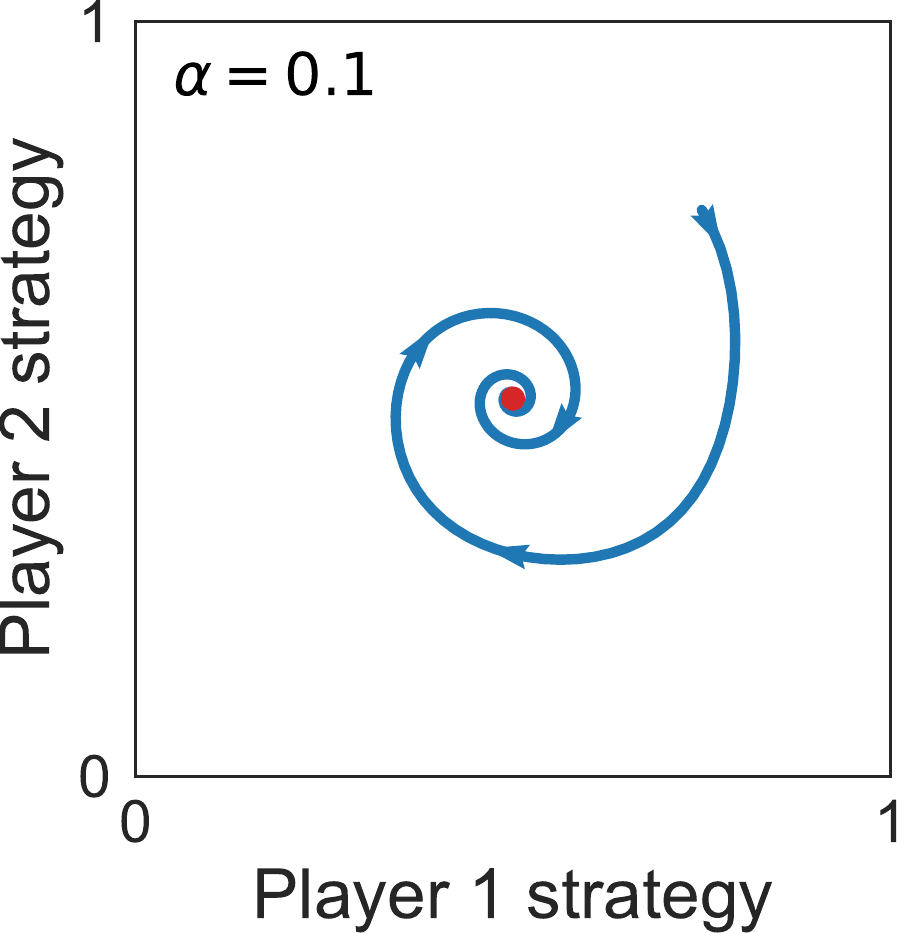}
\includegraphics[width=.49\columnwidth]{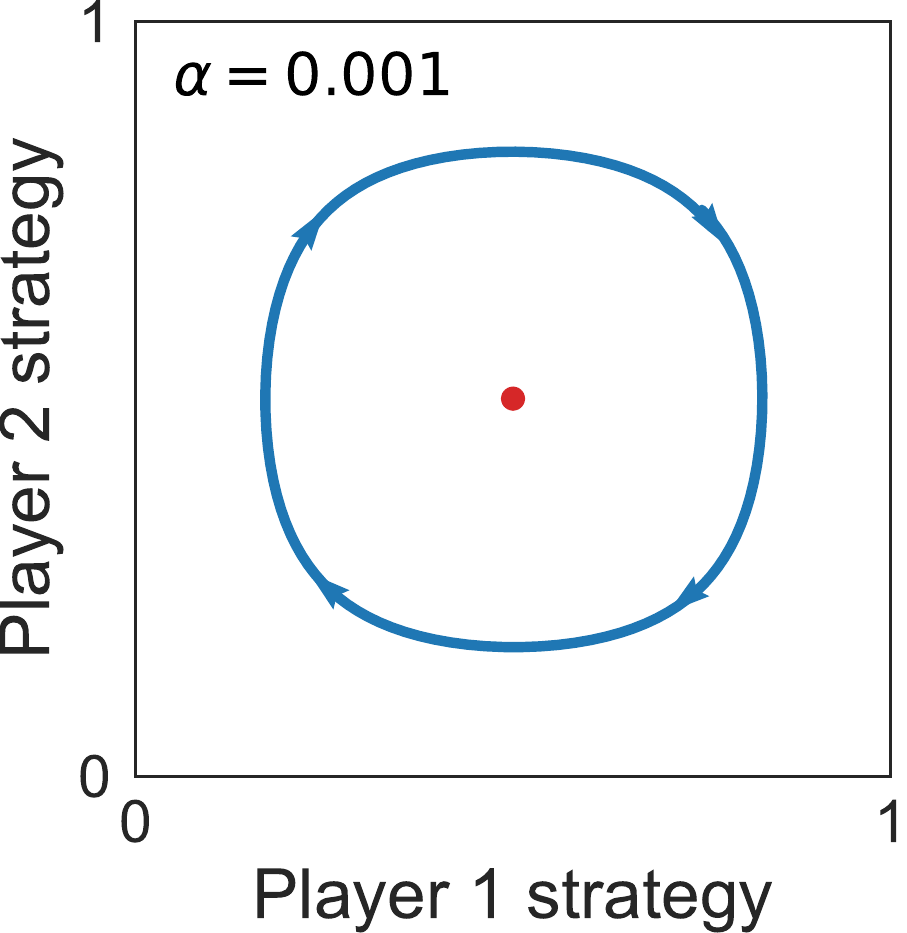}
\caption{Two average trajectories of the Natural Policy Gradient in a zero-sum game, corresponding to two different learning rates. Each point on the trajectories represents a pair $(\pi^1(t),\pi^2(t))$. The average rewards are $\overline{R}_1=((1,-1),(-1,1))$, $\overline{R}_2 = ((-1,1),(1,-1))$. The variance for all the rewards is equal to $\sigma=1$.
The starting point is $(0.75,0.75)$, while the Nash equilibrium is at $(1/2,1/2)$.}\label{fig:2p_NPG}
\end{figure}

The $p$-dimensional $k$-armed bandit introduced here serves as a handy model to unify the description of partitioned policies, multi-state environments, and multi-agent interactions, by mapping them to a disordered system at finite temperature. This can be particularly well illustrated in the case of $p=2$, which can be interpreted as a Matrix Game \cite{vonneumann2007,nash51,berg98,berg99,berg00} between two players, each having its own reward matrix $R_{1(2)}$. It has been shown \cite{galla07}, that replicator dynamics with cooperation pressure $u$ does not converge to all Nash equilibria below a critical value of $u$, unless we deal with a zero-sum game, i.e. $R_1=-R_2^T$. On the other hand, the cooperation pressure, acts in the replicator equation as the mutation term acts in the Langevin approximation of PG. In the case of a zero-sum game, the replicator trajectories can only factorize into a number of converging spirals as shown in the left side of Fig. \ref{fig:2p_NPG}, since Nash equilibria for pure strategies are suppressed for $K\rightarrow\infty$. If, instead, $R_1\neq-R^T_2$, dynamics can converge to pure strategies, but such equilibria have been shown to give birth to a spin glass phase for low values of $u$ \cite{galla07}.
\begin{acknowledgments}
We would like to thank Antonio Celani, Andrea Mazzolini and Enrico Malatesta for the thoughtful discussions and precious insights on the topic.
\end{acknowledgments}

\appendix

\section{Appendix I: Regret bound and optimal learning rate scheduling}

The regret of the Natural Policy Gradient is the difference between the reward obtained by a policy up to time $T$ and the best possible reward one could obtain in the same time. In terms of the $k$-armed bandit problem, it's defined as
\begin{equation}
    \mathcal{R}_T=\max_{a\in\{1,...,k\}}\sum_{t=1}^T R_a^t - \sum_{t=1}^T\sum_{b=1}^k \left<\pi_b^t\right> R_b^t.
    \label{eq:regret}
\end{equation}
One can decompose this expression by introducing the \textit{instantaneous regret} for an arm 
\begin{equation}
    \rho_a^t=R_a^t-\sum_{b=1}^k \pi_b^t R_b^t.
    \label{eq:instant}
\end{equation}
The overall regret for that specific arm will then simply be $\mathcal{R}_{T,a}=\sum_{t=1}^T \left<\rho_a^t\right>$, and therefore the total regret of the policy is the maximum of this quantity over all arms $\mathcal{R}_T=\max_{a\in\{1,...,k\}}\mathcal{R}_{T,a}$. We will consider the rewards to be independent stochastic variables, the only constraint being that they are bounded $R_a^t\in\left[0,R_M\right]$. Nonetheless, the result holds true also for correlated outcomes, non-stationary environments, and, the “unluckiest” configuration that one can imagine.

It\^{o}'s lemma states that if $X_t$ is an It\^{o} drift-diffusion process satisfying the diffusion equation
\[
    dX_{t}=u _{t}dt + \sqrt{2D_{t}}dW_{t},
\]
then any twice-differentiable function $f(X)$ can be expanded to the first order in time following
\[
    df=\left( u_{t}\frac {\partial f}{\partial x} + D _{t}{\frac {\partial ^{2}f}{\partial x^{2}}}\right)dt+\sqrt{2D_t}{\frac {\partial f}{\partial x}}\,dW_{t} + o\left(dt^2\right).
\]
We will apply it to the average \textit{log-policy} $\left<\log{\pi_a}\right>$, expanding it to the form
\begin{equation}
\begin{aligned}
    \frac{d}{dt}\left<\log{\pi_a^t}\right> &= \left< \frac{u_a^t}{\pi_a^t}\right> + \left< \frac{D_{a,a}^t}{(\pi_a^t)^2}\right> = 
    \\
    & \alpha_t \langle \rho_a^t \rangle + \frac{\alpha_t^2}{2} \left( \langle (\rho_a^t)^2 \rangle - \sum_b (R_b^t)^2 (1 - \langle \pi_b^t \rangle) \right)
\end{aligned}
\end{equation}
and by making use of the fact that $\left< (\rho_a^t)^2\right>\geq 0$ and the rewards are bounded $R_b^t\leq R_M \;\forall b,t$, we can write the inequality 
\begin{equation}
    \left<\rho_a^t\right> \leq \frac{1}{\alpha_t}\frac{d}{dt} \left<\log{\pi_a^t}\right> + \frac{\alpha_t}{2}(k-1)R_M^2.
\end{equation}
We can now bound the single-arm regret using the latter equation:
\begin{equation}
\begin{aligned}
    \mathcal{R}_{a,T}\simeq \int_0^T dt \left<\rho_a^t\right> \leq & \left(\frac{\left<\log{\pi_a^T}\right>}{\alpha_T} - \frac{\left<\log{\pi_a^0}\right>}{\alpha_0}\right) + 
    \\
    &\frac{R_M^2}{2}(k-1)\int_0^T dt \alpha_t.
\end{aligned}
\end{equation}
Where we have discarded negative terms. For any final probability distribution $\pi_a^T$, its logarithm will be negative and can be discarded leaving the bound unaltered. If we chose a uniform initial distribution $\pi_a^0 = 1/k \; \forall a$ and assume that $\alpha_T\leq\alpha_0$, we can rewrite the inequality substituting the latter:
\begin{equation}
    \mathcal{R}_{a,T}\leq \frac{\log k}{\alpha_T} + \frac{R_M^2}{2}(k-1) \int_0^T dt \alpha_t.
\end{equation}
As we can see, the choice of scheduling function will influence the regret. 

A convenient functional choice is $\alpha_t=A/\sqrt{t}$. In this way, both contribution are equally weighted ad the expression can be rewritten as
\begin{equation}
    \mathcal{R}_{a,T} \leq \left( \frac{\log k}{A} +R_M^2(k-1)A\right) \sqrt{T}.
\end{equation}
The function $\alpha=A/\sqrt{T}$ can be refined specifying the coefficient $A$ so that the bound is minimised. It's easy to see that such value is $A = \sqrt{\log{k}/(k-1)}/R_M$. Substituting this term, one finds the bound for the regret and the best scheduling of the learning rate for minimising this bound:

\begin{equation}
\mathcal{R}_T \le 2 R_M \sqrt{(k-1) \log k \; T} \hspace{1cm} \alpha_t = \frac{1}{R_M} \sqrt{\frac{\log k}{(k-1) \; t}}
\label{eq:regret_bound}
\end{equation}

\section{Appendix II: The effect of the second-order expansion of the diffusion coefficient}

\begin{figure} [h!]
    \begin{center}
    \includegraphics[width=1.0\columnwidth]{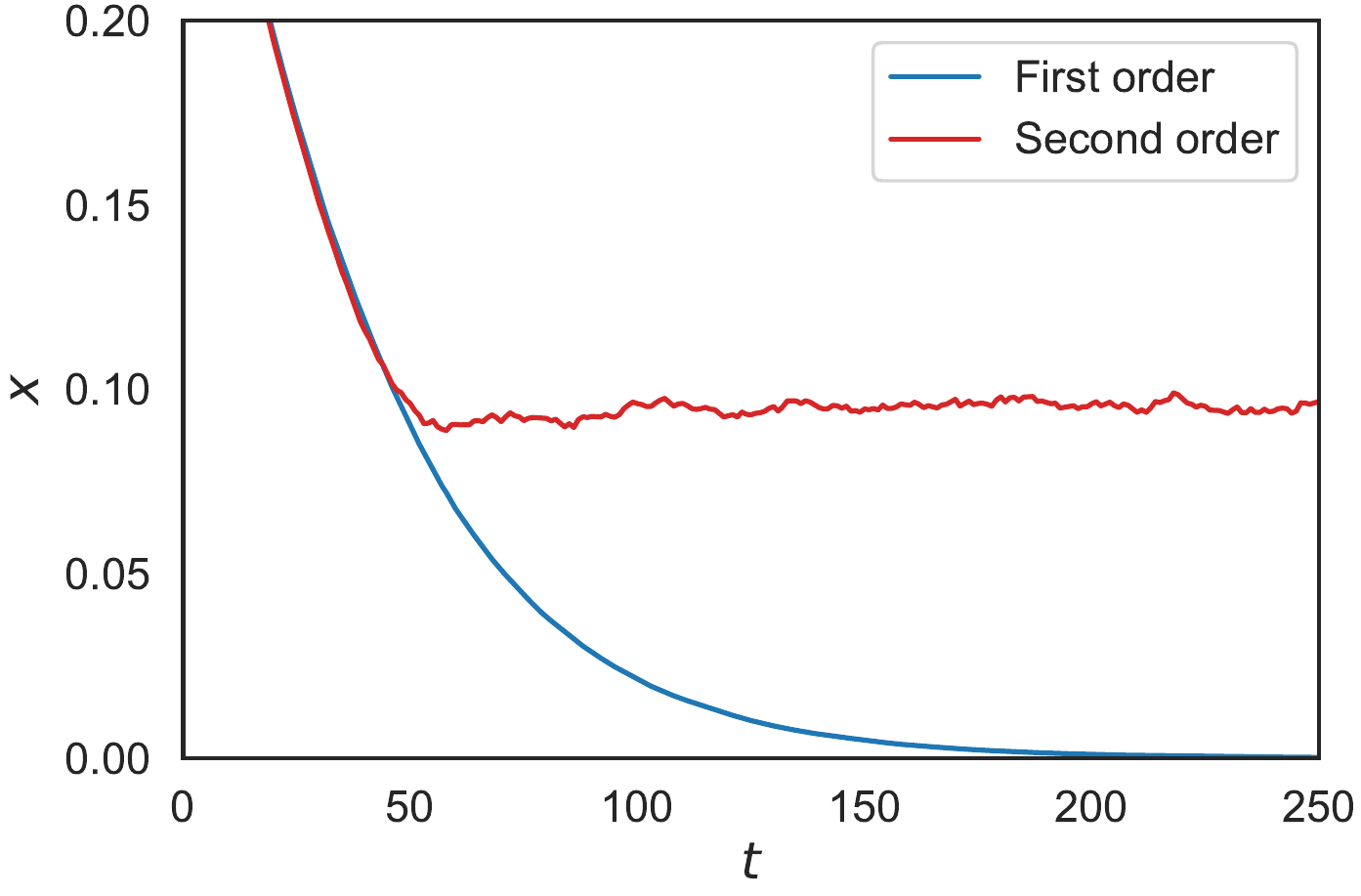}
    \caption{Comparison between average trajectories of the action probability $x$ for the 2-armed bandit updated according to the Langevin dynamics. While the blue curve incorporates only the diffusion coefficient obtained by first-order expansion in $\alpha$, the red one includes also the second-order $\alpha$. }
    \label{sup:diffs}
    \end{center}
\end{figure}

\bibliography{biblio}

\pagebreak
\widetext
\begin{center}
\textbf{\large Supplementary material: Gradient dynamics in reinforcement learning}
\end{center}
\setcounter{equation}{0}
\setcounter{figure}{0}
\setcounter{table}{0}
\setcounter{page}{1}
\makeatletter
\renewcommand{\theequation}{S\arabic{equation}}
\renewcommand{\thefigure}{S\arabic{figure}}
\renewcommand{\bibnumfmt}[1]{[S#1]}
\renewcommand{\citenumfont}[1]{S#1}



\newcommand{\Var}{\mathrm{Var}}

    \section{Detailed derivation of drift-diffusion \\ coefficients}
    \subsection{Two actions}
    In our setting, an agent has a discrete set of actions $A$, which he performs with a certain probability measure $\pi$ at each turn, receiving a payoff $J(\pi)$ by the ambient he is immersed in.
    Our task is to describe the dynamics of the learning algorithm in the space of its parameters. Having only two actions $A_1$ and $A_2$, performed with  probabilities $\pi_1$ and $\pi_2$ respectively, we can use the normalization constraint $\sum_{a=1}^{k} \pi_a = 1$ to reduce the problem to a one-dimensional motion described by the variable $x \equiv \pi_1 = 1-\pi_2$. We will suppose that the two actions at each turn yield stochastic rewards i.i.d. form normal distributions $R_1\sim q_1\equiv\mathcal{N}(r_1,\sigma_1)$ and $R_2\sim q_2\equiv \mathcal{N}(r_2,\sigma_2)$. Consequently, the payoff for a strategy $x$ is equal to
    \begin{equation}
        J(x) = xR_1 + (1-x)R_2.
    \end{equation}
    This dynamic is driven by a gradient ascent $\dot{x} \sim \nabla J(x)$, which reinforces (hence the framework name) the better performing action. Such ascent can not be freely implemented on a compact space as $[0,1]\ni x$. This can be circumvented by the use of a parameter $\theta\in \mathbb{R}$, on which we can freely perform the ascent $\dot{\theta}\sim \nabla J(\theta)$, mapped to $[0,1]$ by the compactification
    \begin{equation}
        x=\frac{1}{1+e^{-\theta}}.
    \end{equation}
    The concrete form we will choose for the ascent is:
    \begin{equation}
        \dot{\theta} =\alpha F^{-1}\nabla J(\theta) = E_\pi[\alpha F^{-1} R\nabla\ln \pi],
    \end{equation}
    where $\alpha$ is the so-called learning rate $F$ and is used to account for the uneven paste yield by the chosen parametrization. It is equal to the Fisher information metric $F=E_\pi[\partial_\theta \ln \pi\partial_\theta \ln \pi]$. This constitutes the algorithm known in literature as natural policy gradient (NPG). At each time step, the algorithm will feel a ``local" gradient, based only on the rewards $R_t$, and its trajectory will thus fluctuate following the stochasticity of the rewards:
    \begin{equation}
        \dot{\theta}_{(t)} = \alpha F_{(t)}^{-1} R_{(t)}\nabla\ln \pi_{(t)}.
    \end{equation}
    
\begin{figure}
    \begin{center}
    \includegraphics[width=0.6\columnwidth]{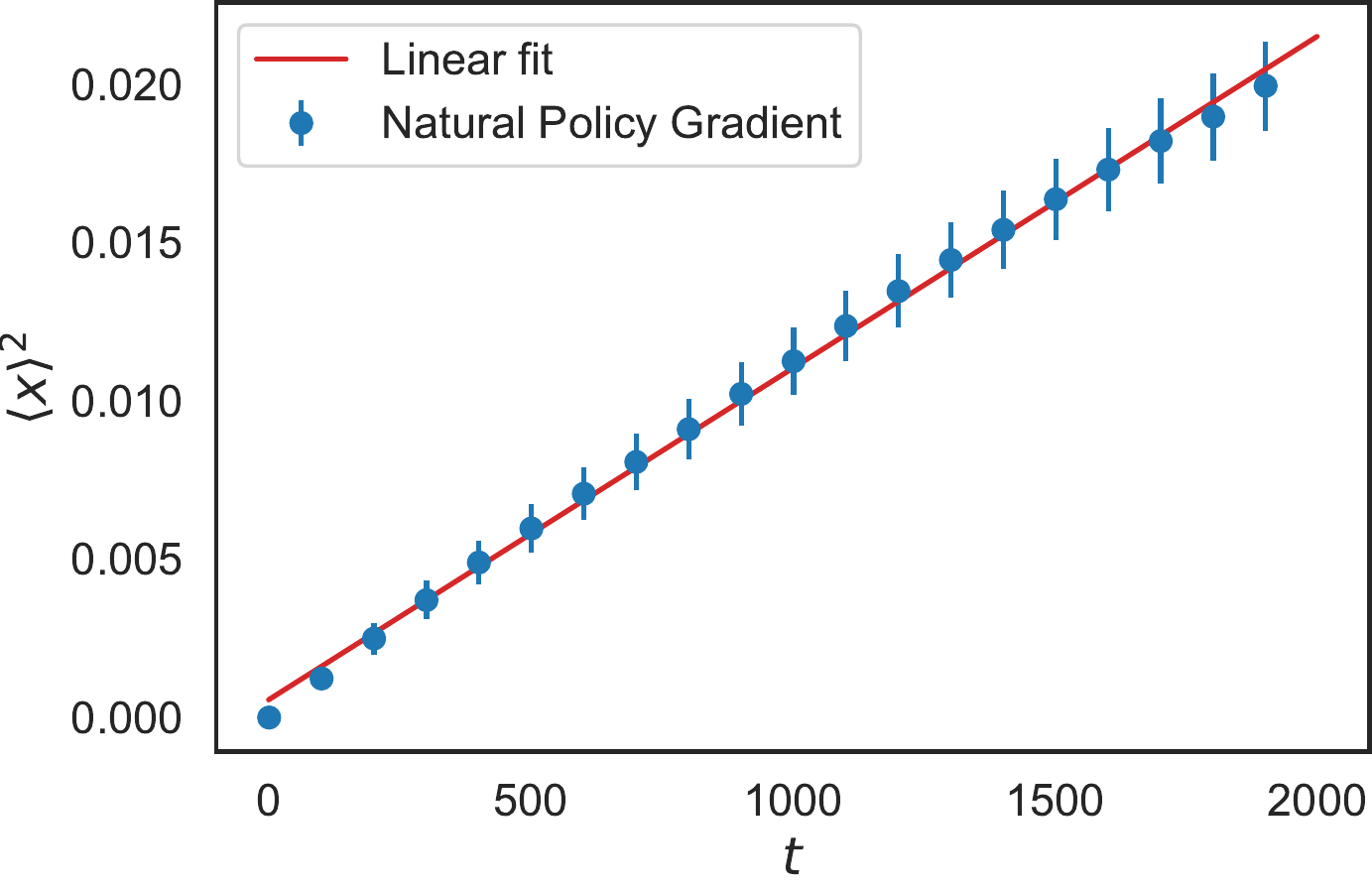}
    \caption{Diffusion of the action probability $x$ for a 2-armed bandit updated according to the natural policy gradient algorithm. The rewards are distributed as $\mathcal{N}(0,0.1)$ and the initial probability is $x_0=0.5$. The linear relation shows that the probability diffuses according to a Brownian motion.}
    \label{sup:fit_std}
    \end{center}
    \end{figure}

    The standard approach for the description of stochastic dynamics is that of a Langevin equation:
    \begin{equation}
    \frac{dx}{dt} = u(x) + \sqrt{2D(x)}\cdot \eta(t),
    \end{equation}\label{sup:lang}
    where $\eta(t)$ is white Gaussian noise with zero mean and correlation $E_t[\eta(\tau)\eta(\tau')] = \delta(\tau-\tau')$. Approximating stochastic fluctuations by Gaussian noise is not always accurate. For instance, the noise of neural networks' stochastic gradient descent is often described by heavy-tailed distributions, which appears to be a crucial characteristic for ensuring convergence to flat minima. Nonetheless, our reinforcement learning setting supposes Gaussian noise in the rewards, which translates to a canonical Brownian motion of the policy gradient, as shown in Fig.\ref{sup:fit_std}.

    In order to obtain the coefficients $u(x)$ and $D(x)$, we will expand $x_t\equiv x(\theta_t)$ supposing a slow learning, i.e. $d\theta\ll 1$, given by $\alpha\ll 1$:
    \begin{equation}
        dx = \frac{dx}{d\theta}d\theta + \frac{1}{2}\frac{d^2x}{d\theta^2} d\theta^2 + o(\alpha^2).
    \end{equation}
    The drift coefficient is then found by taking the average
    \begin{equation}
        u(x) = E_{t}[\dot{x}_t|x_t] = E_{q}[E_\pi[\dot{x}_t|x_t]].
    \end{equation}
    All the handy relations we will use later on are:
    \begin{equation}
        \frac{dx}{d\theta} = x(1-x) ; \quad \frac{d^2x}{d\theta^2} = x(1-x)(1-2x)
    \end{equation}
    \begin{equation}
        F=\sum_\pi\left(\frac{\partial\ln \pi}{\partial\theta}\right)^2 \pi = x(1-x)
    \end{equation}
    \begin{equation*}
        \begin{split}
            \dot{\theta}_t = \alpha F^{-1}\frac{\partial}{\partial\theta} \ln{\pi} \cdot R 
            = \alpha F^{-1}\left( \mathbf{1}_1 R_1 \frac{\partial \ln{x}}{\partial\theta} + \mathbf{1}_2 R_2 \frac{\partial \ln{(1-x)}}{\partial\theta} \right)
        \end{split}
    \end{equation*}
    \begin{equation}
        \dot{\theta}_t = \alpha \left( \mathbf{1}_1 \frac{R_1}{x} - \mathbf{1}_2 \frac{R_2}{1-x} \right)
    \end{equation}
    \begin{equation}
        \dot{\theta_t}^2 = \alpha^2 \left( \mathbf{1}_1 \frac{R^2_1}{x^2} + \mathbf{1}_2 \frac{R^2_2}{(1-x)^2} \right) + \dots
    \end{equation}
    We omitted terms that become irrelevant after the averaging of indicator functions $\mathbf{1}_1 = 1$ with probability $x$ (i.e. $E_\pi[\mathbf{1}_1] = x$), $\mathbf{1}_2 = 1$ with probability $1-x$ (i.e. $E_\pi[\mathbf{1}_2] = 1-x$), and $E_\pi[\mathbf{1}_1\mathbf{1}_2] = 0$, in order to express the probability density in terms of discrete variables. Now, supposing $E_q[R_a] = r_a , \; E_q[R_a^2] = \sigma_a^2 + r_a^2$, we obtain
    \begin{equation}
        \begin{split}
            u(x) &= E_q[E_\pi[\frac{dx}{d\theta}\dot{\theta}_t + \frac{1}{2}\frac{d^2x}{d\theta^2}\dot{\theta}_t^2|x_t]]  
            \\
            & = \alpha x(1-x)(r_2-r_1) + \frac{\alpha^2}{2}(1-2x)[(1-x)(\sigma_1^2 + r_1^2) + x(\sigma_2^2 + r_2^2)].
            \end{split}
    \end{equation}

Analogously, for the diffusion term:
    \begin{equation}
        \begin{split}
            2D(x) &= \Var_t[\frac{dx}{d\theta}\dot{\theta}_t + \frac{1}{2}\frac{d^2x}{d\theta^2}\dot{\theta}_t^2|x_t]  
            \\ 
            &=E_q[E_\pi[(\frac{dx}{d\theta}\dot{\theta}_t + \frac{1}{2}\frac{d^2x}{d\theta^2}\dot{\theta}_t^2)^2|x_t] - E_\pi[\frac{dx}{d\theta}\dot{\theta}_t + \frac{1}{2}\frac{d^2x}{d\theta^2}\dot{\theta}_t^2|x_t]^2].
        \end{split}
    \end{equation}
    The first part of $2D(x)$ is:
    \begin{equation}
        \begin{split}
            & E_q[E_\pi[(\frac{dx}{d\theta}\dot{\theta}_t)^2|x_t] - E_\pi[\frac{dx}{d\theta}\dot{\theta}_t |x_t]^2] =
            \\
            & =  \alpha^2 x^2(1-x)^2 \left( \frac{\sigma_1^2 + r_1^2}{x} + \frac{\sigma_2^2 + r_2^2}{1-x}\right) -
            \alpha^2 x^2(1-x)^2(r_2-r_1)^2 
            \\
            & = \alpha^2x(1-x)\left[ (1-x)\sigma_1^2 + x \sigma_2^2 +((1-x)r_1 + xr_2 )^2\right].
        \end{split}\label{1order_diff}
    \end{equation}

    The second part is:
    \begin{equation}
             E_q[E_\pi[2\frac{dx}{d\theta}\dot{\theta}_t\cdot\frac{1}{2}\frac{d^2x}{d\theta^2}\dot{\theta}_t^2|x_t] - 2E_\pi[\frac{dx}{d\theta}\dot{\theta}_t |x_t] E_\pi[\frac{d^2x}{d\theta^2}\dot{\theta}_t^2|x_t] \approx 0.
    \end{equation}
    This term is proportional to $x(1-x)(1-2x)$, and, thus, almost everywhere equal to zero.
    
    The third part is:
    \begin{equation}
        \begin{split}
            & E_q[E_\pi[(\frac{1}{2}\frac{d^2x}{d\theta^2}\dot{\theta}_t^2)^2|x_t] - E_\pi[\frac{1}{2}\frac{d^2x}{d\theta^2}\dot{\theta}_t^2|x_t]^2]
            \\
            & = \frac{\alpha^4}{4}(1-2x)^2[(1-x)^2\frac{E_q[R_1^4]-xE_q[R_2]^2}{x} +x^2 \frac{E_q[R_2^4]-(1-x)E_q[R_2]^2}{1-x} 
            \\
            & +2x(1-x)E_q[R_1]E_q[R_2] ] 
            \\
            & \approx \frac{\alpha^4}{4}(1-2x)^2  [(1-x)^2\frac{(3-x)(\sigma_1^2+r_1^2)^2-2r_1^4}{x} +x^2 \frac{(2+x)(\sigma_2^2+r_2^2)^2-2r_1^4}{1-x} ].
        \end{split}\label{2order_diff}
    \end{equation}
    
    We clearly see how the dynamic is driven towards a deterministic solution ($x=1$ or $x=0$), by the term of $u(x)$ proportional to $\alpha$, whilst it is repelled from it by the term proportional to $\alpha^2$. The diffusion analogously affected by $\alpha$: whilst its first power governs diffusion in the bulk of the probability simplex ($x\approx 1/2$), $\alpha^2$ is responsible for the diffusion next to solutions. A remarkable behavior is underlined by this expansion: the diffusion, although dumped by the square of the learning rate, augments in the vicinity of a solution. The difference between Langevin dynamics with first and second-order diffusion coefficients is shown in Fig.\ref{sup:diffs}. The second-order term of $D$ is the main dissimilarity between the Kimura equation and NPG's Langevin approximation. It is clear that it acts repelling the probability $x$ from boundaries, whereas the first-order term vanishes.

\begin{figure}
    \begin{center}
    \includegraphics[width=0.6\columnwidth]{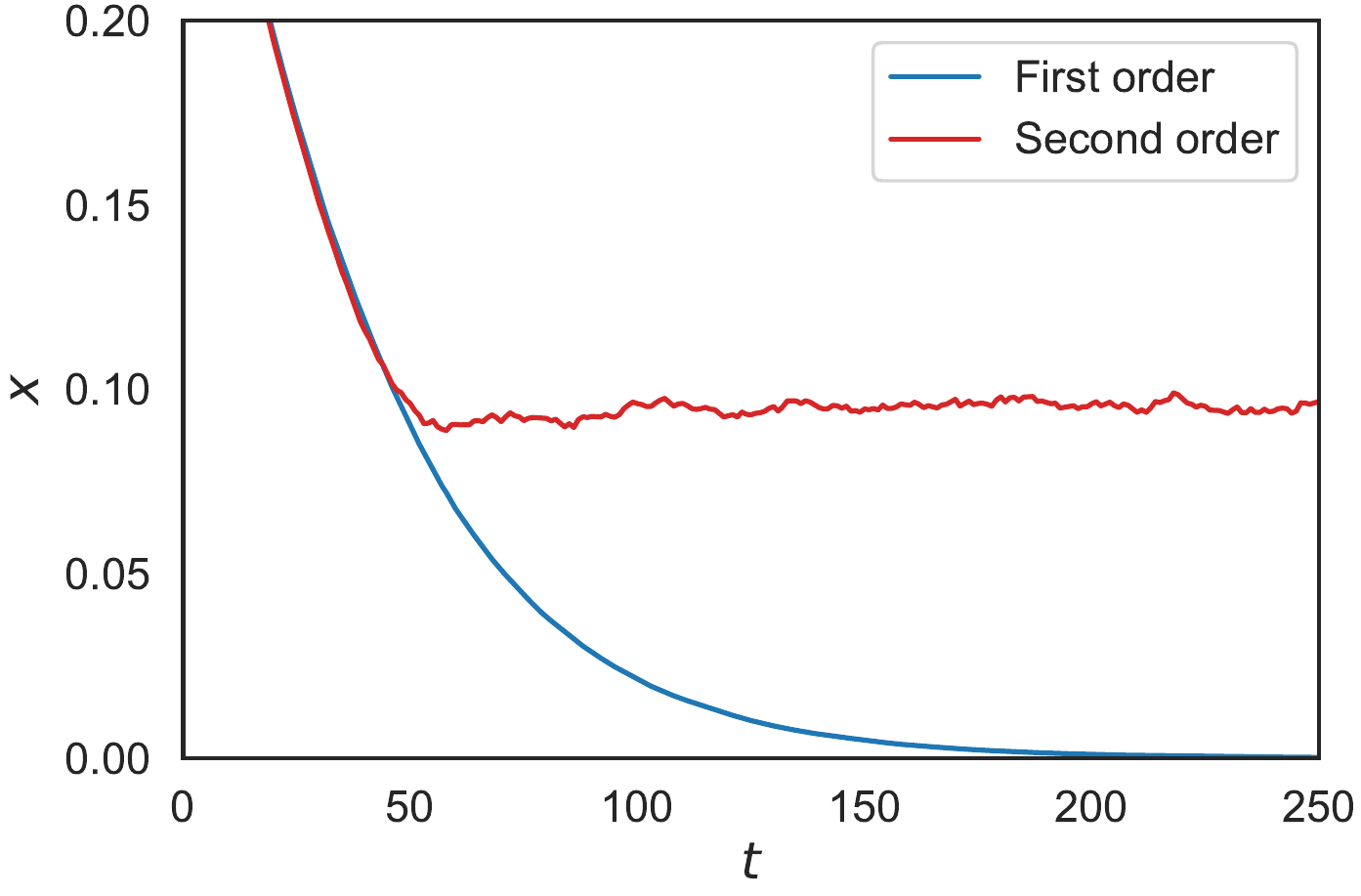}
    \caption{Comparison between average trajectories of the action probability $x$ for the 2-armed bandit updated according to the Langevin dynamics. While the blue curve incorporates only the diffusion coefficient obtained by first-order expansion (\ref{1order_diff}), the red one includes also the second-order (\ref{2order_diff}). }
    \label{sup:diffs}
    \end{center}
    \end{figure}

\subsection{K actions}

    When dealing with $k$ independent actions, the dynamics will follow a \\ $k$-dimensional drift-diffusion motion that can be described in the Langevin or It\^{o} form. This time we will use the latter, but for our purposes they are equivalent.
    \begin{equation}
     d\pi_a = u_a(\pi)dt + \sum_{ab}\sigma_{ab}(\pi) dW_b \quad,\label{sup:multidim}
    \end{equation}
    in which $W_b$ are the components of a $k$-dimensional Wiener process. 

    The generalization to a set composed of $k$ independent actions is achieved assigning $k$ parameters $\theta_{a\in\{1,...,k\}}$ to their respective probabilities by the mappings
    \begin{equation}
        \pi_a(\boldsymbol{\theta})=\frac{e^{\theta_a}}{\sum_{b=1}^k e^{\theta_b}}\quad.
    \end{equation}
    
    Fluctuations in the probability space $W_{\pi}$ are not uncorrelated. Correlations arise from the constraint $\sum_{a=1}^k \pi_a =1$, which yields $\Cov_\pi[W_{\pi a},W_{\pi b}]=\rho_{ab}$. In order to describe the process in terms of uncorrelated Gaussian noises, we define noise in the parameters' $\theta$ space, which is mapped to the probability space via $W_{\pi a} =\sum_b \frac{\partial\pi_a}{\partial \theta_b}W_b$, where $\Cov_\pi[W_a,W_b]=\delta_{ab}$.
    
    It naturally follows, that the probabilities can be expanded as
    \begin{equation}
        d \pi_a = \sum_b \partial_b \pi_a d \theta_b +  \sum_{bc} \partial_b \partial_c \pi_a d \theta_b d \theta_c + o(\alpha^2)    
    \end{equation}
    As before, the gradient ascent will be given by
    \begin{equation}
        \dot{\boldsymbol{\theta}}_{(t)} = \alpha \; F_{(t)}^{-1} \; R_{(t)}  \; \boldsymbol{\nabla}\log\pi(\boldsymbol{\theta}_{(t)}),
    \end{equation}
    where $(F)_{ab} = E_\pi \left[ \partial_{a} \ln \pi(\boldsymbol{\theta}) \partial_{b} \ln \pi(\boldsymbol{\theta}) \right]= \pi_a (\delta_{ab} - \pi_b)$. It is now clear why the metric $g_{ij}=F_{ij}$ is useful: the gradient is a covariant vector field on a smooth statistical manifold, i.e. a Riemannian manifold each of whose points is a probability distribution. Our manifold is defined by the map $\pi(\boldsymbol{\theta})$ and the local covariant basis of tangent vectors is defined by $\boldsymbol{e}_i=\partial\log\pi/\partial\theta_i$. Having this in mind, it is easy to see how we need to account for the controvariance of $d\theta^i$, while $\partial_{i}\log\pi$ is covariant. We simply can't equate two such vectors, since they behave in opposite ways under the curvature's effects. Of course, with the help of the metric $g_{ij}=\boldsymbol{e}_i\boldsymbol{e}_j = F_{ij}$, we can transform covariant into contravariant vectors, thus obtaining $d\theta^i\propto g^{ij}\partial_j\log\pi$, or $d\boldsymbol{\theta} \propto F^{-1} \boldsymbol{\nabla} \log\pi$.
    
    Useful relations for later on derivations include
    \begin{equation}
    \partial_{b} \pi_a = \pi_a(\delta_{ab} - \pi_b)
    \end{equation}
    \begin{equation}
    \partial_{b}\partial_{c} \pi_a = \pi_a \left[ (\delta_{ab} - \pi_b)(\delta_{ac} - \pi_c) -\pi_b (\delta_{bc} - \pi_c)\right]
    \end{equation}
    \begin{equation*}
        \dot{\theta}_{a} = \alpha \sum_{b,c=1}^k F^{-1}_{ab} R_c (\delta_{bc} - \pi_b) \mathbf{1}_c ,
    \end{equation*}
    \begin{equation}
     \dot{\theta}_{a} = \alpha \frac{R_{a}}{\pi_a} \mathbf{1}_a
    \end{equation}
    \begin{equation}
       \dot{\theta}_b \dot{\theta}_c = \alpha^2\frac{R_b R_c}{\pi_b \pi_c} \mathbf{1}_a\mathbf{1}_b = \alpha^2\frac{R_b R_c}{\pi_b \pi_c} \delta_{bc}\pi_b
    \end{equation}
    The drift coefficient $u_a(\pi)$ is found by averaging:
    \begin{equation}
     \begin{split}
        u_a &= E_q[E_\pi[\dot{\pi}_a|\pi]] 
        \\
        & = E_q[E_\pi[\sum_b \partial_b\pi_a \dot{\theta}_b|\pi]] + E_q[E_\pi[\sum_{bc}\partial_c\partial_b\pi_a \dot{\theta}_b \dot{\theta}_c|\pi]] =u_a^1 + u_a^2
     \end{split}
    \end{equation}
    \begin{equation}
         u_a^1 = \alpha\pi_a \left( r_a-\sum_b r_b\pi_b\right) 
    \end{equation}
    \begin{equation}
        \begin{split}
       u_a^2 &= E_q[E_\pi[\frac{\alpha^2}{2}\sum_{bc}\partial_b \partial_c \pi_a \frac{R_bR_c}{\pi_b\pi_c}\mathbf{1}_b\mathbf{1}_c ]]
       \\
       &=E_q[\frac{\alpha^2}{2}\sum_{bc}\partial_c  \partial_b  \pi_a \frac{R_bR_c\pi_c}{\pi_c\pi_d} \delta_{cd} ]
       \\
       &= \frac{\alpha^2}{2}\sum_c \pi_a \left[ (\delta_{ac} - \pi_c)(\delta_{ac} - \pi_c) -\pi_c (\delta_{cc} - \pi_c)\right]\frac{E_q[R_c^2]}{\pi_c}
       \\
        &= \frac{\alpha^2}{2}\pi_a\sum_c  \left( \delta_{ac}^2 -2\pi_c\delta_{ac} + 2\pi_c^2 -\pi_c\delta_{cc} \right)\frac{E_q[R_c^2]}{\pi_c}
        \end{split}
    \end{equation}
    \begin{equation}
        u^2_a = \frac{\alpha^2}{2}\left(E_q[R_a^2] (1-2\pi_a) -\sum_{b}^k E_q[R_b^2](1-2\pi_b)\pi_a \right).
    \end{equation}
    Supposing $E_q=[R_a^2]=\sigma^2_a$, be obtain
    \begin{equation}
         u_a^2 =
         \frac{\alpha^2}{2}\left(\sigma_a^2 (1-\pi_a)(1-2\pi_a) -\sum_{b\neq a}^k \sigma^2_b(1-2\pi_b)\pi_a \right).
    \end{equation}
    
    The diffusion coefficient is found upon taking the covariance of the increments:
    \begin{equation}
    \begin{split}
          \frac{1}{2}\Cov_t[\dot{\pi_a},\dot{\pi}_b] &= \frac{1}{2}\Cov_q[\Cov_\pi[\dot{\pi}_a,\dot{\pi}_b]] 
          \\
         &=\frac{1}{2}\Cov_q[\Cov_\pi[\sum_\alpha \sigma_{a\alpha}W_\alpha,\sum_\beta \sigma_{b\beta}W_\beta]] =
         \frac{1}{2}\boldsymbol{\sigma}\boldsymbol{\sigma}^T = \boldsymbol{D}
    \end{split}
    \end{equation}
    Expanding $\dot{\pi}_a$ up to the first order in $\alpha$, we find
    \begin{equation}
    \begin{split}
       2D_{ab} &= E_q[\Cov_\pi[\sum_{c}\partial_c  \pi_a d\theta_c , \sum_{d}\partial_d  \pi_b d\theta_d ]]
       \\
        &= E_q[E_\pi[\sum_{cd}\partial_c  \pi_a \partial_d  \pi_b d\theta_d d\theta_c ]]
        \\
        &= E_q[E_\pi[\sum_{cd}\partial_c  \pi_a \partial_d  \pi_b \frac{\mathbf{1}_c\mathbf{1}_d}{\pi_c\pi_d} ]]
        \\
        &=E_q[\sum_{cd}\partial_c  \pi_a \partial_d  \pi_b \frac{\delta_{cd}\pi_c}{\pi_c\pi_d} ]
        \\
        &=\sum_{c}\pi_a \pi_b (\delta_{ac}-\pi_c)(\delta_{bc}-\pi_c) \frac{E_q[R_c^2]}{\pi_c}
        \\
        &= \pi_a \pi_b \sum_{c} (\delta_{ac}\delta_{bc}-\pi_c\delta_{ac} - \pi_c\delta_{bc} +\pi_c^2) \frac{E_q[R_c^2]}{\pi_c}.
    \end{split}
    \end{equation}
    Supposing $E_q=[R_a^2]=\sigma^2_a$, be obtain
    \begin{equation}
       2D_{ab} = \pi_a\pi_b\left(\delta_{ab}\frac{\sigma_a^2}{\pi_a} + \sum_{c\neq a,b} \pi_c \sigma_c^2 -(1-\pi_a)\sigma_a^2 - (1-\pi_b)\sigma_b^2  \right).
    \end{equation}
    
    To summarize, we found the coefficients governing the drift-diffusion motion:
\begin{equation}
\begin{aligned}
     u_a = & \alpha\pi_a \Bigg( r_a-\sum_b r_b\pi_b\Bigg) + \\
     & \frac{\alpha^2}{2}\Bigg(\sigma_a^2 (1-\pi_a)(1-2\pi_a) -\sum_{b\neq a} \sigma^2_b(1-2\pi_b)\pi_a \Bigg),\\
 &\\
       D_{ab} = & \frac{\alpha^2}{8} \pi_a\pi_b\Bigg(\delta_{ab}\frac{\sigma_a^2}{\pi_a} + \sum_{c\neq a,b} \pi_c \sigma_c^2
       \\
       &  \qquad \qquad -(1-\pi_a)\sigma_a^2 - (1-\pi_b)\sigma_b^2  \Bigg).
\end{aligned}
\label{eq:coef_simply}
\end{equation}
It is now clear that the drift is driven towards the maximum by a replicator dynamic, and away from pure strategies by the mutation term. In addition, the diffusion term, here expanded up to the second order in alpha, scatters the trajectory proportionally to the rewards' variances.

\section{Details of the mean-field method}

The $p$-dimensional $k$-armed bandit can be viewed as an agent that at each time step is performing $p$ independent actions, each one chosen from a unique set of $K$ possible actions. An example of such agent is the robotic hand with its five fingers flexing independently of one another. Despite actions being independent, they yield an overall result that can be optimized with a Policy Gradient. If each finger can be flexed to $K$ different angles, then we can assign a probability $\pi_{i}$ to each angle $i\in\{1,\dots,K\}$. Each overall configuration will then yield a reward based on $p$ distribution probabilities:
\begin{equation}
\begin{split}
    J=&E_{\pi_1,\pi_2,\ldots,\pi_p}[R_{1,2,\ldots,p}]
    \\
    =&\sum_{\pi_1,\pi_2,\ldots,\pi_p=1}^K R_{i_1,i_2,\ldots,i_p}  \pi_{i_1}\cdot\pi_{i_2}\cdot\ldots\cdot\pi_{i_p}.
\end{split}\label{sup:pdim}
\end{equation}
If we suppose that the rewards are normally distributed around their averages $\mathcal{N}(\overline{R}_{i_1 i_2 \ldots i_p},\sigma_{i_1 i_2 \ldots i_p})$, we can consider the system described by the Hamiltonian $H\equiv -\overline{J}$ obtained substituting mean rewards in (\ref{sup:pdim}). Its temperature $T(\sigma)$ is defined by the specific learning algorithm, and for a policy gradient is proportional to the diffusion coefficient of the Langevin dynamics (\ref{sup:multidim}). One can clearly see how this total reward has multiple local minima, generated by the quenched disorder of the coefficients $\overline{R}_{i_1,i_2,\ldots,i_p}$. In other words, it corresponds to the Hamiltonian of a planar $p$-spin model, i.e. with constraints $\sum_{i_a}^K \pi_{i_a}=1, \quad \pi_{i_a}>0,\quad \forall a\in\{1,\ldots,p\}$.
The analogy with a spin glass becomes clear if we imagine a magnet in which interatomic forces are extremely weak. Heating it up, atoms start to oscillate before spins. This will subsequently affect spins, since their interaction depends on their distance, resulting in an effective temperature for them, probed by the intensity of their Brownian motion. In order to analyze the structure of minima arising in this problem, one can study its free energy: 
\begin{equation}
F=-\frac{1}{\beta}\ln\sum_{\{\pi\}}\exp\{\beta J[\pi]\} = -\frac{1}{\beta}\ln Z,
\end{equation}
where $\sum_{\{\pi\}}$ means summing over all possible values of $(\pi_{i_1}\pi_{i_2}\ldots,\pi_{i_p})$, each one corresponding to one configuration, generating a particular reward $J[\pi]$. Including a temperature via the parameter $\beta=1/T$, lets us explore the landscape of $J$, since for high temperatures all configurations are weighted equally in the resulting free energy, while lowering $T$ permits us to see how local minima arise. Of course, every problem has its own energetic landscape defined by its own constants $R_{i_1,i_2,\ldots,i_p}$. In order to study average properties, we will average over the disorder by considering it a self-averaging quantity, meaning that we will take each instance to be drawn from a normal distribution $\mathcal{N}(0,\frac{1}{K})$. To render the problem tractable, we will use the mean field method named replica trick: $\overline{\ln Z} = \lim_{n\to 0} \frac{1}{n} \ln \overline{Z^n}$. Each of the $n$ replica of the system will have a mean partition function
\begin{equation}\label{sup:part1}
\begin{split}
\overline{Z} &= \int_0^\infty\prod_{i=1}^K d^p\pi_i \, \delta^p(\sum_i\pi_i - K)  
\int_{-\infty}^{+\infty} \prod_{i_1,\ldots,i_p} dR_{i_1\ldots i_p} \\
\\
& \times \exp{\left[-R^2_{i_1\ldots i_p}K^p +\beta R_{i_1\ldots i_p}\pi_{i_1}\ldots\pi_{i_p}\right]},
\end{split}
\end{equation}
where the integration is performed over $p$ sets of $d\pi_1d\pi_2\dots d\pi_K$, with $p$ constraints $\sum\pi=K$ enforced by $p$ delta functions. Once we multiply $n$ copies of the system, we integrate over the disorder variables, obtaining
\begin{equation}\label{sup:part2}
    \begin{split}
    \overline{Z^n} =& \int_0^\infty \prod_{a=1}^n\prod_{i=1}^K d^p\pi_i^a \, \delta^p(\sum_i\pi_i^a - K)   \exp{\left[\frac{\beta^2}{4K^{p-1}}\sum^n_{a,b}\pi^a_{i_1}\pi^b_{i_1}\dots\pi_{i_p}^a \pi_{i_p}^b\right]}
    \\
    =& \int D\pi^a_i \exp{\left[\frac{\beta^2}{4K^{p-1}}\sum^n_{a,b}\left(\sum_i^K\pi^a_i\pi^b_i\right)^p\right]}.
    \end{split}
\end{equation}
In order to render this expression tractable, we will make an ansatz on the form of $Q_{ab}=\sum_{i=1}^K \pi_i^a\pi_i^b$. To enforce it, we insert the identity $1=\int\delta(Q_{ab}-\sum_i\pi_i^a\pi_i^b)dQ_{ab}$. Delta functions can be dealt with by passing to their Fourier transform:
\begin{equation}
\begin{split}
    & \delta(\sum_i\pi_i^a-K)=\int_{-\infty}^{+\infty}\exp\left[-ik\left(\sum_{i,a}\pi_i^a\xi^a-K\sum_a\xi^a\right)\right]\prod_a^n \frac{d\xi^a}{\sqrt{2\pi}},
    \\
    & \delta(Q_{ab}-\sum_i\pi_i^a\pi_i^b)=\int_{-\infty}^{+\infty}\exp\left[-ik\left(\sum_{a,b}Q_{ab}\Lambda_{ab}-\sum_{a,b}\Lambda_{ab}\pi^a\pi^b\right)\right]\prod_{a,b}^n \frac{d\Lambda_{ab}^a}{\sqrt{2\pi}}.
\end{split}
\end{equation}
Absorbing $-ik$ into $\xi^a$ and $\Lambda_{ab}$, we finally arrive at the complete expression for the replicated partition function
\begin{equation}
    \begin{split}
        \overline{Z^n}=&\int_{-\infty}^{+\infty}\prod_{a,b}^n\prod_i^K dQ_{ab}d\Lambda_{ab}d\xi^a\int_0^{+\infty}\prod_a^n\prod_i^Kd\pi_i^a \times
        \\
        &\times\exp\left(\frac{\beta^2 K}{4} \sum_{ab}Q^p_{ab} + K\sum_{ab}Q_{ab}\Lambda_{ab} - \sum_i\sum_{ab}\Lambda_{ab}\pi^a_i\pi^b_i \right.\\
   &\left. -\sum_{ia}\xi^a\pi^a_i + K\sum_a\xi_a \right)
    \end{split}
\end{equation}
For large $K\rightarrow\infty$, the integral is dominated by saddle point. We can now impose various ansatzes to the form of $Q_{ab},\Lambda_{ab}$ and $\xi^a$, the simplest of which is the replica-symmetric:
\begin{align*}
       Q_{aa} = q_0 & \qquad \Lambda_{aa} = \lambda_0 & \xi^a = \xi & \qquad \forall a, \\
       Q_{ab} = q_1 & \qquad \Lambda_{ab} = \lambda_1 & \forall a>b, \addtocounter{equation}{1}\tag{\theequation}
\end{align*}
which yields a system of equations for the saddle point
\begin{equation}\label{sad1}
\begin{split}
    S= & \text{extr}_{q_0,q_1,\lambda_0,\lambda_1,\xi}\left[ \frac{\beta^2}{4T^2} \sum_{ab}(q_1+(q_0-q_1)\delta_{ab})^p + q_0\lambda_0  \right. \\
    & + q_1\lambda_1 + \xi +\ln \left. \int  \prod d\pi e^{- (\xi \sum \pi +\sum\lambda_{ab}\pi^a\pi^b)}\right]
\end{split}
\end{equation}

Such picture gets modified if we shift our frame of reference from the probabilities $\pi$ to the parameters $\theta$. The latter are performing an ascent along the gradient of the reward $\dot{\boldsymbol{\theta}}\sim\alpha\boldsymbol{\nabla}J$, which speed is governed by the slope of the surface $J(\boldsymbol{\pi})$ and the learning rate $\alpha$. From their perspective, they are affected by a resulting slope $\dot{\boldsymbol{\theta}}_t\sim R_{1,2,\ldots,p} \boldsymbol{\nabla}\ln(\pi_1\pi_2\ldots\pi_p)^\alpha = R_{1,2,\ldots,p} \boldsymbol{\nabla}\ln\phi_1\phi_2\ldots\phi_p$, corresponding to a gradient over the reward $J_\alpha=E_{\phi^{1/\alpha}}[R]$. Performing a change of variables $\pi=\phi^{1/\alpha}$ in \ref{sup:part1}, we obtain 
\begin{equation}\label{sup:part3}
    \begin{split}
     \int D\phi^a_i \exp{\left[\frac{\beta^2}{4K^{p-1}}\sum^n_{a,b}\left(\sum_i^K\left(\phi^a_i\phi^b_i\right)^\frac{1}{\alpha}\right)^p\right]},
    \end{split}
\end{equation}
where $D\phi_i^a$ contains the Jacobian and the constraints. In the neighborhood of a pure strategy, i.e. when $\phi\rightarrow(1,0,\dots,0)$, one can expand $(\sum_i\phi_i^a\phi_i^b)^{1/\alpha}\approx\sum_i(\phi^a_i\phi_i^b)^{1/\alpha}$, since all non-diagonal terms are suppressed, thus obtaining a saddle point equation analogous to (\ref{sad1}). This time, the equation for $Q_{ab}$ will read
\begin{equation}
    0=\frac{p}{4T^2\alpha}Q_{ab}^{\frac{p}{\alpha}-1} + \Lambda_{ab}
\end{equation}
which means that we operate as if we had a new temperature $T\rightarrow \sqrt{\alpha}T$.

\end{document}